\documentclass[authoryear,final,3p,times,twocolumn]{elsarticle}

\usepackage{amssymb}
\usepackage{amsmath}
\usepackage{booktabs} 
\usepackage{graphicx}

\usepackage{subfigure}
\usepackage{xspace}
\usepackage[colorlinks=true,     linkcolor=blue,     anchorcolor=blue,
citecolor=blue,            filecolor=blue,            menucolor=blue,
urlcolor=blue]{hyperref}

\newcommand{\Msun}{\ensuremath{{M}_\odot}}
\newcommand{\diff}{\ensuremath{\mathrm{d}}} 
\newcommand{\deriv}[2]{\ensuremath{\frac{\diff {#1}}{\diff {#2}}}}
\newcommand{\eps}{\ensuremath{\varepsilon}}
\newcommand{\mean}[1]{\ensuremath{\left\langle #1 \right\rangle}}
\newcommand{\abs}[1]{\left\lvert#1\right\rvert}

\newcommand{\vect}[1] {\ensuremath{\mathbf{#1}}}

\newcommand{\hMpc}[1]{\ensuremath{{#1}\,h^{-1} \mathrm{Mpc}}}
\newcommand{\Mpch}[1]{\ensuremath{{#1}\,h\, \mathrm{Mpc}^{-1}}}

\newcommand{\hMsun}[1]{\ensuremath{{#1}\,h^{-1} \Msun}}
\newcommand{\zi}[1]{\ensuremath{z_\text{ini} = {#1}}}
\newcommand{\gad}{\texttt{Gadget}}
\newcommand{\go}{\texttt{GOTPM}}

\newcommand{\modif}[1]{{#1}}

\journal{New Astronomy}

\begin{document}

\begin{frontmatter}



\title{Effects of the initial conditions on cosmological $N$-body simulations}


\author[kias]{Benjamin~L'Huillier}
\ead{lhuillier@kias.re.kr}
\author[kias]{Changbom~Park}
\ead{cbp@kias.re.kr}
\author[kias,cac]{Juhan~Kim\corref{cor1}} 
\ead{kjhan@kias.re.kr}
\cortext[cor1]{Corresponding author}

\address[kias]{
School of  Physics,  Korea  Institute for  Advanced
  Study, 85 Hoegi-ro, Dongdaemun-gu, Seoul 130-722, Korea}
\address[cac]{
Center for Advanced Computation, Korea Institute for Advanced
  Study,  85 Hoegi-ro, Dongdaemun-gu, Seoul 130-722,  Korea}

\begin{abstract}
  Cosmology  is entering  an era  of  percent level  precision due  to
  current large observational surveys.  
  This precision  in observation is now demanding  more accuracy from
  numerical methods and cosmological simulations. 
  In this paper, we study  the accuracy of $N$-body numerical simulations
  and their dependence on changes in the initial conditions and in the
  simulation algorithms.  
  For  this  purpose,  we  use  a  series  of  cosmological  $N$-body
  simulations with varying initial conditions.  
  We  test  the  influence  of  the  initial  conditions,  namely  the
  pre-initial  configuration  (preIC),  the  order of  the  Lagrangian
  perturbation theory (LPT), and the initial redshift ($z_\text{ini}$), on the
  statistics  associated  with  the  large  scale  structures  of  the
  universe such as the halo mass function, the density power spectrum,
  and the maximal extent of the large scale structures.  
  We  find  that glass  or  grid  pre-initial  conditions give  similar
  results at $z\lesssim 2$.
  However, the initial excess of power in the glass initial conditions
  yields  a  subtle difference  in  the  power  spectra and  the  mass
  function at high redshifts.  
  The  LPT  order used  to  generate  the  initial conditions  of  the
  simulations is found to play a crucial role. 
  First-order  LPT  (1LPT)  simulations  underestimate the  number  of
  massive haloes  with respect to second-order  (2LPT) ones, typically
  by 2\% at \hMsun{10^{14}} for an initial 
  redshift of  23, and the small-scale power  with an underestimation
  of 6\% near the Nyquist frequency for \zi{23}.  
  Larger underestimations are observed for lower starting redshifts.
  Moreover,  at  higher  redshifts,  the  high-mass end  of  the  mass
  function is significantly underestimated in 1LPT simulations.  
  On  the  other hand,  when  the LPT  order  is  fixed, the  starting
  redshift has  a systematic  impact on the  low-mass end of  the halo
  mass function.  
  Lower starting redshifts yield more low-mass haloes.
  Finally, we compare two $N$-body codes, \gad{-3} and \go, and find
   8\% differences  in the power spectrum  at small scales  and in the
   low-mass end of the halo mass function. 

\end{abstract}

\begin{keyword}
Cosmology:  simulations \sep  Cosmology:  large-scale structures  \sep
Method: numerical 

\end{keyword}

\end{frontmatter}

 
\section{Introduction}

Cosmological simulations  have proved to  be one of the  most powerful
tools  for  the  study  of  the non-linear  evolution  of  large-scale
structures (LSS) of the Universe. 
In the hierarchical cold dark matter model with a cosmological constant
($\Lambda$CDM), structure  formation occurs from  bottom up, with
the collapse of primordial density fluctuations forming low-mass haloes
that     later     merge     to     form     more     massive     ones
\citep[e.g.][]{1984Natur.311..517B,   1985ApJ...292..371D},  while  the
rate and  type of mergers  are governed by the  large-scale background
environment \citep[e.g.][]{2009ApJ...700..791H, interactions}.  
The pioneering work  of \citet{1974ApJ...187..425P}, later improved by
more refined approaches, \citep[e.g.][]{2002MNRAS.329...61S}, provides
us  with important  physical insights  on the  evolution of  the large
scale  structures  through  the   study  of  the  halo  mass  function
\citep[e.g.        ][]{2001MNRAS.321..372J,       2006ApJ...646..881W,
  2008ApJ...688..709T}.  

Cosmology has entered an era of percent-level precision thanks to the 
recent   high-resolution  cosmic  microwave   background  observations
\citep{2013ApJS..208...19H,   2013arXiv1303.5062P}  and   huge  galaxy
redshift  surveys   (e.g.   2dF,  \citealt{1999RSPTA.357..105C};  BOSS,
\citealt{2013AJ....145...10D}; DES).  
In  concert  with  observational advancements,  analytic  perturbation
theories have also been refined  and can now describe the quasi-linear
evolution of density fields to higher precision. 
\modif{%
  However,   comparing  the  results   of  numerical   simulations  to
  theoretical expectations is not an easy task.  
  Because of  the non-linear nature of structure  formation, no theory
  can accurately predict  the mass function or the  non-linear part of
  the power spectrum.  
  One has  to rely  on fits to  cosmological simulations to  model the
  halo      mass     function     \citep[e.g.,][]{2002MNRAS.329...61S,
    2006ApJ...646..881W, 2008ApJ...688..709T}  or the non-linear power
  spectrum  \citep{2003MNRAS.341.1311S} to  study  structure formation
  and to confront cosmological models to observations.
}  
In making accurate comparisons between models and observations, it is
now  necessary   to  know  whether  numerical   simulation  can  yield
convergent results, and 
to determine the simulation parameters to produce accurate results.
Several  studies  reported  that  the  results  of  a  simulation  are
sensitive to the choice of the starting redshift 
\citep{2007ApJ...671.1160L,  2009ApJ...698..266K, 2010ApJ...715..104H,
  2013MNRAS.431.1866R}, the order of the Lagrangian 
perturbation theory \citep{1998MNRAS.299.1097S, 
  2006MNRAS.373..369C, 2010MNRAS.403.1353C, 2010MNRAS.403.1859J}, 
or the  initial distribution  of the particles  prior to  applying the
displacement, namely on a  regular lattice (grid), or a glass
configuration \citep{1994astro.ph.10043W, 2007MNRAS.380...93W}. 
\citet{2007ApJ...671.1160L}   claimed   that    the 
root mean square (rms) of  the displacement should be small (typically
0.20  times  the  mean  particle  separation), and  that  the  Nyquist
frequency should be in the linear regime at the starting epoch. 
They also claimed that there  should be at least about 10 expansion
factors between $z_\text{ini}$ and the  first redshift of interest so that the
memory of the initial grid or glass configuration is lost.
\citet{2009ApJ...698..266K} inspected the influence of the starting
redshift  on   the  inner  structure  of   haloes  (triaxiality,  spin
parameter, and concentration), using both 1LPT and 2LPT.
They found that starting redshift  and LPT order have little influence
on the internal   halo   properties. 
Initial conditions generated by 1LPT are known to produce transients,
and  2LPT can  be  used to  reduce  the effects  of these  transients
\citep{2006MNRAS.373..369C}.

The  initial configuration  should satisfy  several  requirements: the
configuration  should be isotropic,  homogeneous, and  should be  in a
state of equilibrium.
The most  popular preICs  are regular lattice  (grid) and  glass, but
several other configurations have also been introduced \citep[e.g.
quaquaversal tiling, ][]{2007ApJ...656..631H}.
Setting  the initial  configuration  of particles  on  a regular  grid
introduces preferred directions.  
Moreover,   warm  and   hot   dark  matter   simulations  have   shown
unphysical features on the scale of the pixel size, suggesting
that  grid preICs  are  not well  suited  when structures  form in  a
top-down way  and the initial  power of density fluctuation  vanishes on
the pixel scale. 
Some authors \citep[e.g.][]{1994astro.ph.10043W}  advocate the use of
more elaborated techniques like glass preICs.  
For this configuration, particles are randomly (uniformly) placed on a
grid, then the set of 
particles is advanced by a  repulsive gravitational force law until it
reaches an equilibrium state where particles feel virtually no forces.  
However, the drawback of this technique  is that reaching the
equilibrium is  computationally expensive,  and one should  test whether
 the configuration is indeed in a state of equilibrium. 
Another problem  is that the  density field started from  glass preICs
contains  spurious clustering  on  small scales,  which  does not  grow,
but should be taken carefully  into account, especially when one wants
to study high-redshift physics. 

\begin{table*}
  \begin{center}
    \caption{\label{tab:sim} 
      Simulations used in this study.  
      Set 1  consists of \go{}  simulations, with the  equivalent \gad{}
      simulation for T2s, T3s and T4s.  T3s and G3 have been run with 4
      different realisations. 
      Set 2 uses a WMAP7 cosmology with different random seeds.
    }
    \begin{tabular}{lrcrllll}
      \toprule
      ID  & $N_\text{p}$  &$L_\mathrm{box}$ &  $z_\text{ini}$ &  LPT &
      PreIC & r.m.s./$\Delta x$ & Comments\\
      & &(\hMpc{}) & & & & &  \\
      \midrule
      Set 1: &  & & & & & & \go{}, WMAP5 cosmology\\
      T2s & 1024$^3$ & 256 & 100 & 2 & mesh & 0.47 & \gad{-3} run: G2s\\
      T3s & \phantom{1}512$^3$ & 256 & 100 & 2 & mesh & 0.26 & 4 realisations;
      \gad{-3} run: G3s\\ 
      T3f & \phantom{1}512$^3$ & 256 & 100 & 1 & mesh & 0.26 & 4 realisations \\
      T4s  & \phantom{1}512$^3$  & 256  & 50  & 2  & mesh  & 0.51  & 4
      realisations; \gad{-3} run: G4s\\ 
      T4f & \phantom{1}512$^3$ & 256 & 50 & 1 & mesh & 0.51 & 4 realisations \\
      T5s & \phantom{1}512$^3$ & 256 & 23 & 2 & mesh & 1.09 & 4 realisations \\
      T5f & \phantom{1}512$^3$ & 256 & 23 & 1 & mesh & 1.09 & 4 realisations\\
      T7s & \phantom{1}512$^3$ & 768 & 100 & 2 & mesh & 0.089 & \\
      T8s & \phantom{1}512$^3$ & 768 & 50 & 2 & mesh & 0.18 & \\
      T9s & \phantom{1}512$^3$ & 768 & 23 & 2 & mesh & 0.38 & \\
      \midrule  
      Set 2:& & & & & & & \gad{-3}, WMAP5 glass/mesh\\
      G3sm & \phantom{1}512$^3$ & 256 & 100 & 2 & mesh &&  \\
      G3sg & \phantom{1}512$^3$ & 256 & 100 & 2 & glass & & \\
      G4sm & \phantom{1}512$^3$ & 256 & 50 & 2 & mesh &  & 4 realisations\\
      G4sg & \phantom{1}512$^3$ & 256 & 50 & 2 & glass & & 4 realisations\\
      \midrule
	      Set 3:& & & & & & & \gad{-3}, WMAP7\\      
      G7s & \phantom{1}512$^3$ & 768 & 50 & 2 & mesh & & \\
      G7f & \phantom{1}512$^3$ & 768 & 50 & 1 & mesh & & \\
      G8s & \phantom{1}512$^3$ & 768 & 23 & 2 & mesh & & \\
      G8f & \phantom{1}512$^3$ & 768 & 23 & 1 & mesh & & \\
      \bottomrule
    \end{tabular}
  \end{center}
\end{table*}

In the  last few years several  authors have begun  to investigate the
possibility of percent  level accuracy of the power  spectrum (PS) and
halo     mass    function     (MF)     in    numerical     simulations
\citep{2010ApJ...715..104H,2013MNRAS.431.1866R}.  
\citet{2010ApJ...715..104H} used  a high $z_\text{ini}$ and 1LPT  rather than a
lower initial redshift with  2LPT, arguing that higher initial
redshift yields convergence  in the mass function at  $z=0$ and is not
computationally expensive. 
However, 
\citet{2013MNRAS.431.1866R}  found that too  high a  starting redshift
($z_\text{ini}  \gtrsim  200$  for   a  mean  particle  separation  of
\hMpc{2}) produces an erroneous  mass function because the amplitude of
the initial fluctuations is too low compared to the level of numerical
noise. 
They claim that 2LPT, with at least 10-50 expansion factors
before the first redshift of interest, should be used.
Moreover,  they studied the  influence of  several parameters  of Tree
codes on the halo mass function, and the power spectrum.
 
\modif{The aim of this paper is to study the effect of changes in the initial
conditions at fixed cosmology on the large-scale structure statistics.  
To do so, we quantify the influence of the preICs, the LPT order,
and the 
starting redshift on large-scale  statistics such as the density power
spectrum, the halo mass function,  and the distribution of the size of
structures.  }
Since we aim to study the effects of the initial conditions on
the  properties  of   large-scale  structures  (LSS),  the  individual
properties of haloes are not studied in this paper. 

Section \ref{sec:sim} presents the set  of simulations we used in this
work. 
Section \ref{sec:stat}  presents the statistics  that we used  for the
analysis. 
Our main  results are shown  in \S~\ref{sec:res}, and  our conclusions
are drawn in \S~\ref{sec:ccl}. 
\ref{sec:Pk} aims to find the best estimation of the power spectrum.

\section{Simulations}
\label{sec:sim}

\subsection{Initial Conditions}

The comoving positions of particles in the initial conditions is computed via
\begin{equation}
  \label{eq:zeldov}
  \vect x(\tau) = \vect{q} + \vect{\Psi}(\vect q,\tau),
\end{equation}
 \noindent where $\diff\tau =  \diff a(t)/a(t)$ is the conformal time,
 $\vect x$  is the Eulerian  (after displacement) position,  $\vect q$
 the Lagrangian 
 (initial) position, 
and $\vect \Psi$ the displacement field.
Here, $\vect{x}$ obeys the equation
\begin{equation}
\deriv{^2 \vect{x}}{\tau^2} + \mathcal{H}(\tau) \deriv{\vect{x}}\tau =
- \vect{\nabla \phi}, 
\end{equation}
where $\phi$ is the potential field, and $\mathcal{H}=aH$.
The displacement field $\vect{\Psi}$ is given to the second order by
\begin{equation}
\vect{\Psi}  (\vect{q},\tau)   =  -D_1(a)  \vect{\nabla}_q  \phi^{(1)}
(\vect{q}) + D_2(a) \vect{\nabla}_q \phi^{(2)} (\vect{q}),
\end{equation}
where the  $D_i$ are  the first- and  second-order growth  factors, and
$\phi^{(i)}$ are the first- and second-order potentials.
The      first       order      approximation,      which      assumes
$\vect{\nabla}\phi^{(2)}=0$, is called the Zel'dovich approximation.  

The initial conditions for our first set of simulations were generated
by the IC generator included in  \go{}, which is able to generate both
1LPT   and   2LPT   ICs,   using   a   CAMB   \footnote{available   at
  \url{http://camb.info/}} input power spectrum at $z=0$.  
The ICs for the second and third sets of simulations were generated using the
2LPTic code described in \citet{2006MNRAS.373..369C}.

 \subsection{Sets of simulations}

Our simulations used a WMAP5 cosmology (sets 1 and 2) with
$(\Omega_\text{b},  \Omega_\text{m},\Omega_\Lambda,h,\sigma_8,n_s)$  =
(0.044,  0.26,  0.74,  0.72, 0.79,  0.96),  and  set  3 used  a  WMAP7
cosmology (0.0455, 0.272, 0.728, 0.702, 0.807, 0.961).  
The simulation parameters are summarised in Table~\ref{tab:sim}.
Our  reference   set  of   simulations  uses  the   \texttt{GOTPM}  code
\citep{2004NewA....9..111D,2005ApJ...633...11P}.
\go{}   has   long  been   used   to   run  cosmological   simulations
\citep{,2005ApJ...633....1P}, including the recent Horizon runs 
\citep{2009ApJ...701.1547K, 2011JKAS...44..217K}.
This   TreePM   code    is   the   merger   of   the    PM   code   of
\citet{1990MNRAS.242P..59P}      and     the     Tree      code     of
\citet{1996NewA....1..133D}.  
A second set of simulations was run using the TreePM-SPH
\texttt{Gadget-3}   code,   an   enhanced   version  of   the   public
\texttt{Gadget-2} code \citep{2005MNRAS.364.1105S}.

Hereafter, we will  refer to the simulations using  their names in the
table,  and  to  groups  of  simulations with  their  common  features
(number of  particles $N$, order  of LPT, preIC, initial  redshift, or
box size).
Our  choice  of  the  softening  parameter  ($\eps_\text{soft}  =  0.1
\bar{d}$)    is   slightly    above   the    recommended    value   of
\citet{2013MNRAS.431.1866R}.  
The size of  the PM grid was  set equal to the number  of particles in
each dimension, in both \gad{} and \go{} runs.

Rather  than  running  several  identical simulations  with  different
random seeds, we  chose to use the same random  seed, thus allowing us
to  observe  the  variation   of  the  initial  parameters  free  from
statistical fluctuations. 
However,  this  introduces a  bias  while  comparing simulations  with
different resolutions  due to the  random fluctuations in  the initial
conditions.  
In order to  alleviate this issue, we ran three  additional T3, T4, T5
1LPT and 2LPT simulations, and the corresponding G3, G4, and G5 simulations that we used for
the comparison of \gad{} and \go.

\section{Statistics}
\label{sec:stat}

\subsection{Power spectrum}
The matter density power spectrum is defined as 
\begin{equation}
P(k) = \frac 1 V \mean{\abs{\delta(\vect{k})}^2},
\end{equation}
where   $V$   is  the   volume   on   which   it  is   computed,   and
$\delta(\vect{k})$ is the Fourier transform\footnote{In our convention 
  the Fourier transform of a function $f$ is $\hat{f}(\vect{k}) = \int
  f(\vect{x})\exp(i  \vect{k}\cdot\vect{x})\diff^3\vect{x}$,  and
  ${f}(\vect{x})        =         \frac        1        {(2\pi)^3}\int
  \hat{f}(\vect{k})\exp(-i\vect{k}\cdot\vect{x})\diff^3\vect{k}$}
of the overdensity field defined as $\delta(\vect{x}) =
\rho(\vect{x})/\bar{\rho}-1$.  

\citet{2005ApJ...620..559J}  found a formula  to remove  the windowing
effects from the PS that arise due to mass assignment schemes. 
They      showed      that       the      raw      power      spectrum
$\mean{\abs{\delta^f(\vect{k})}^2}$ measured from the simulation using
a fast Fourier  transform (FFT) is related to  the real power spectrum
$P(\vect{k})$ via 
\begin{align}
\mean{\abs{\delta^f(\vect{k})}^2}                                     &=
\sum_{\vect{n}\in\mathbb{Z}^3}\abs{W(\vect{k}    +    2   k_\mathrm{N}
  \vect{n})}^2P(\vect{k} + 2 k_\mathrm{N} \vect{n}) \nonumber\\
&
+  \frac  1
N_\text{p} \abs{W(\vect{k} + 2 k_\mathrm{N} \vect{n})}^2, 
\end{align}
where $k_\text{N} = {N_\text{p}\pi}/L$  is the Nyquist frequency,  the
first  term in  the  second hand  of  the equation  is  the effect  of
aliasing and 
the  convolution  by  the  window  function  $W$  owing  to  the  mass
assignment 
 scheme, and the second term is the Poisson shot noise.
Another  way to  deal with  this aliasing  effect is  to  optimise the
choice of the window function \citep{2008ApJ...687..738C}.
\citet{2009MNRAS.393..511C}  introduced a Taylor-Fourier  transform to
accurately measure the power spectrum.

The estimation of the power spectrum is discussed in~\ref{sec:Pk}.  
We use  a triangular shaped  cloud (TSC) mass assignment  scheme, using
$N_\text{grid}^3  = 2^3  N_\text{p}^3$, and  corrected  for pixelation
effect. 
However we do not correct for shot noise nor aliasing.
Finally, we  use a  cloud-in-cell (CIC) scheme  to bin the  modes into
linearly spaced bins of $2\pi/L$.

\subsection{Mass function}

Haloes  were  detected  at  $z=0$  using  a  friends-of-friends  (FOF)
algorithm.  
FOF links together particles with  a separation smaller than $b$ times
the mean interparticle distance, where $b$ is the so-called linking
length parameter.  
Usually, a  value of  $b=0.2$ is  chosen since it  has been  shown to
reproduce fairly accurately the abundance of virialised haloes
\citep[eg,][]{1994MNRAS.271..676L}.
We computed the mass function of FOF haloes with more than or equal to
20  particles  for  each  simulation,  yielding a  mass  threshold  of
\hMsun{1.8\times 10^{11}}  for the simulations  with $512^3$ particles
and a box size of \hMpc{256} (e.g. T3, T4, T5).  
\modif{
While 20 particles  is definitely not enough to  obtain robust results
on the internal properties of low-mass haloes, since we are interested
in the differences in the mass  function, this issue should not be too
important here.
}

\modif{
The mass of a halo is not well defined in simulations. 
We use here the  FOF mass, that is the sum of  the masses of particles
composing the halo.  
We used the canonical value  $b=0.2$ for the linking length parameter,
which yields haloes with isodensity contours of $\approx 75$ times the
critical density \citep{2009ApJ...692..217L}.  
The boundaries  of the  FOF haloes can  have complex geometry  but are
well-defined  in  terms of  the  local  density  (they are  isodensity
surfaces). 
The FOF  halo finding algorithm is  the most commonly  used method for
estimation  of  the halo  mass  function  in cosmological  simulations
\citep[eg][]{2001MNRAS.321..372J,2006ApJ...646..881W}.  
Its results are known to differ from Spherical Overdensity halo masses
\citep[cf][for example]{2009ApJ...692..217L, 2011ApJS..195....4M}. 
We  chose  to use  FOF  haloes because  SO  by  definition only  finds
spherical haloes, while FOF can deal with more complex geometries.  
}
 
\subsection{Size of the large scale structures}
\label{sec:lss}
In addition  to the  power spectrum and  the halo mass  function, which
have been  used in previous  studies, we also studied  connectivity of
the LSS.  
We identified the LSS in our simulations
using the method proposed by \citet{2012ApJ...759L...7P}.
We applied the FOF algorithm to the halo catalogue, in order to group
them. 
This is a robust and simulation-independent way to define
the large scale structures. 
We varied $b$ from 0.2 to 0.8, with a step of 0.025, to find the value
$b_\text{max}$ that maximises the number of structures.
A low  value of $b$, corresponding  to a high  density threshold, will
produce many small  objects, while a large value  (low threshold) will
produce few very large ones. 
Interestingly, for all these  simulations,  we  found   a  value  of
$b_\text{max}$ between  0.5 and 0.55.
To  detect the  subtle  dependence of  LSS  on the  choice of  initial
conditions, we fixed the linking length to $b=0.55$ to identify the LSS
in all simulations. 
The mean  halo separations are  {3.92} and  \hMpc{4.02}  for T3s
and T7s.
For  each of  these large  scale structures,  we computed  the maximal
extent $l_\text{max}$, which is the maximal distance between each pair
of haloes within the  structure, and study the cumulative distribution
of the maximal extent. 
In the higher resolution simulations, T2s and G2s, the analysis was
made using only  haloes that can be resolved in the lower resolution
simulations, namely,  $M_\text{min} = \hMsun{1.80\times  10^{11}}$, so
that we effectively study the same kind of objects in different simulations. 

\section{Results}

\label{sec:res}

In  this section  we present  the  main results  of our  study on  the
effects of the  preIC, the order of perturbation  theory, the starting
redshift, and the effect of the $N$-body code.  
We use statistics such as  the power spectrum, the halo mass function,
and the distribution of the maximal extent of the LSS. 
Our results should be taken  with caution because the magnitude of the
effects presented below depends on the resolution of the simulations.  
In most cases, we are using simulations with mean particle separations
of 0.5 and \hMpc{1.5}.

\subsection{Pre-initial configuration}

\begin{figure*}[t!] 
  \begin{center} 
    \includegraphics[width=\textwidth]{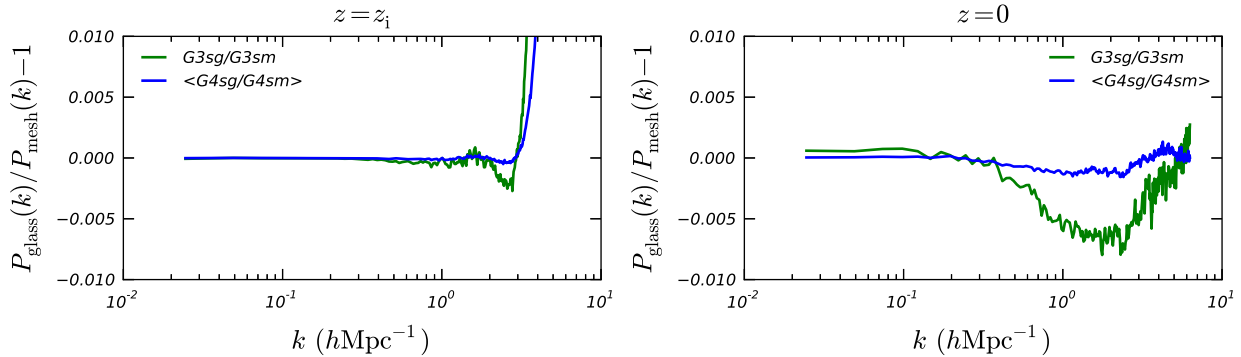}
    \caption{\label{fig:Pk_glass} Effects  of the preICs  on the power
      spectrum. Shown  are the power  spectra of the  glass simulation
      divided by those of the grid ones for G3sg and G3sm (green), and
      the average of the four G4sg and G4sm.  
    }
  \end{center} 
\end{figure*}

The first  variable that  we investigated is  that of  the pre-initial
configuration,  which took the  form of  either a  glass or  a regular
mesh.  
Figure  \ref{fig:Pk_glass}  shows  the  power  spectra  of  the  glass
simulations divided  by that of the corresponding  mesh simulations, at
initial (left) and final  (right) redshifts, for the G3sg/G3sm (green)
and G4sg/G4sm (blue) simulations. 
In  the  G4sg  and  G4sm  cases,  we  took  the  average  over  the  4
realisations. 
At the initial  epoch, the agreement is about  $0.1\% $ until $k\approx
\Mpch{3}$, and the  power of glass-preICs steeply rises  for very high
$k$. 
The power spectrum  is misrepresented up to a few  pixel scales in the
case of the glass preIC.
At $z=0$, the agreement is better than our desired level of 1\% at all
scales. 
The initial excess of power on  small scales has been overtaken by the
power produced by the gravitational clustering.
We note that the difference between glass and mesh preICs is larger in
the   G3sg/G3sm   ($z_\text{ini}    =   100$)   than   the   G4sg/G4sm
($z_\text{ini}= 50$) case.  
This is  still true when  we only consider  the first run of  G4sg and
G4sm, which shares the same random seed as the G3sg/G3sm run.

\begin{figure}[h!]
  \begin{center}
    
      \includegraphics[width=222pt]{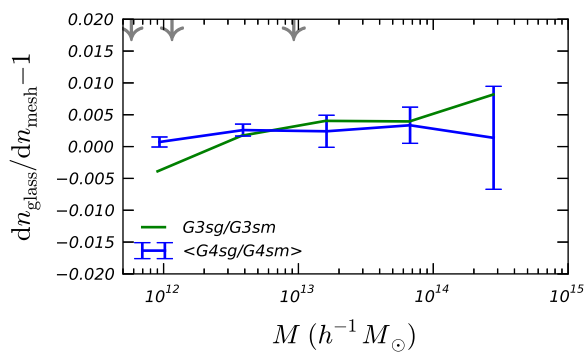}
       \caption{\label{fig:mf_glass} Effects of  the  preICs on  the
      halo mass function.  
      Shown are the mass functions   of  the glass simulation divided  by those of  the grid ones
      for G3sg and G3sm (green), and  the average of the four G4sg and
      G4sm (blue). 
      The grey arrows show the mass of 64, 128, and 1024
      particles. }
  \end{center}
\end{figure}

\begin{figure}[t!]
  \begin{center}
      \includegraphics[width=222pt]{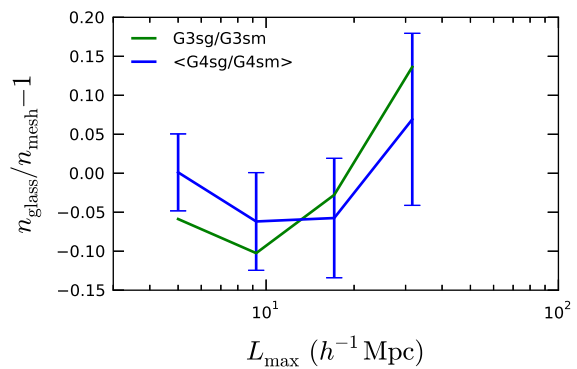}
       \caption{\label{fig:lss_glass}Effects of  the  preICs on  the
      size distribution of the large-scale structures, same legend as Fig.~\ref{fig:mf_glass}. }
    
     \end{center}
\end{figure}

Figure  \ref{fig:mf_glass} shows  the mass  function at  $z=0$  of the
glass  versus   grid  preIC   simulations,  averaged  over   the  four
realisations in the case of G4sg/G4sm.  
The  error bar  is the  standard deviation  in each  bin,  showing the
variations among the four runs. 
The grey arrows show the mass of 64, 128, and 1024 particles. 
At all  masses, glass and grid  simulations agree to better than  1\%.
Figure \ref{fig:lss_glass} shows the effects of the preICs on the size
distribution of LSS. 
It shows the ratio  of the cumulative distributions $n(>L_\text{max})$
of the glass and grid simulations.
The error bars are now much  larger, up to 15\%, due to the relatively
small simulation  box size, but  the glass and grid  simulations agree
within these error bars. 
 
\begin{figure}[t!]
   \begin{center}

     \includegraphics[width=222pt]{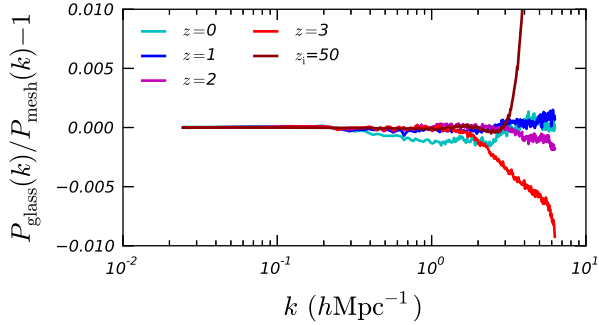}
     \caption{       \label{fig:Pk_glass_z} 
       Redshift evolution of  ratio of the glass to  grid power spectra
      averaged  over the four
       G4sm and G4sg simulations.  
       }
   \end{center}
 \end{figure}

\begin{figure}[t!]
   \begin{center}
     \includegraphics[width=222pt]{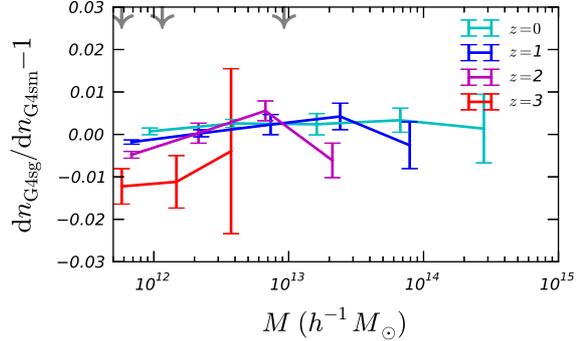}
     \caption{\label{fig:mf_glass_z} 
       Redshift evolution of  ratio of the glass to  grid halo mass function averaged  over the four
       G4sm and G4sg simulations.  
       The grey arrows show the mass of 64, 128, and 1024 particles.
     }
      
   \end{center}
\end{figure}

Even though the initial power  spectra are different, the final results
seem to be consistent with each other. 
One  may then  wonder when  these differences  are washed  out  by the
non-linear gravitational evolution.  
 Figure \ref{fig:Pk_glass_z}  shows the evolution of the  ratio of the
 glass to  grid power  spectra, averaged over  the four G4sm  and G4sg
 simulations.  
At $z=3$, when the scale factor increased by a factor of about 13, the
difference is already smaller than 1\%, and becomes smaller than 0.2\%
at $z=2$.  
Interestingly, at  $z=3$, the glass  simulation show less  small scale
power than the grid case.  
This reflects the  fact that non-linear evolution occurs  later in the
glass case.
Figure  \ref{fig:mf_glass_z}  shows  the   ratios  of  the  halo  mass
functions of the four G4sm and G4sg simulations at different redshifts.
Up to $z=2$, the mass functions agree within a few percent. 
At $z=3$, the number  density of low-mass haloes ($M<\hMsun{10^{13}}$)
is underpredicted  in the glass  simulations with respect to  the grid
one, which is consistent with the power spectrum results.

\subsection{Order of perturbation theory}

\begin{figure*}[t!] 
  \begin{center}
    \includegraphics[width=\textwidth]{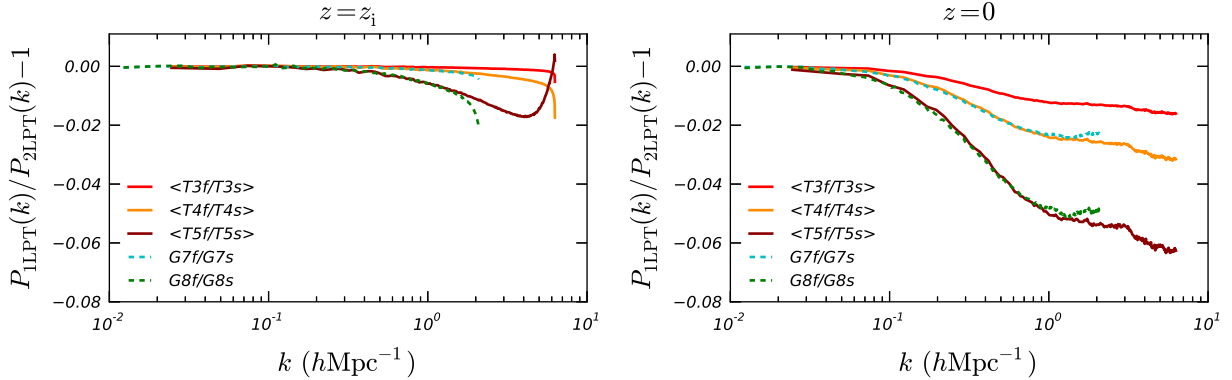}
    \caption{\label{fig:Pk_lpt}  Influence  of the  LPT  order on  the
      power spectrum at $z=z_\text{ini}$ (left) and $z=0$ (right).  
Shown  are the ratios  of the  1LPT to  2LPT simulations  for starting
redshifts of  100 (T3), 50 (G7 and  T4), and 23 (G8  and T5), averaged
over four realisations in the case of T3, T4, and T5. 
    }
  \end{center}
\end{figure*}

Figure \ref{fig:Pk_lpt} shows the effects of the LPT order on the power
spectrum at the initial (left) and final (right) epochs, for different
simulations.
The initial redshifts are 100 for T3  and G7, 50 for T4 and G8, and 23
for T5.
The T (G)simulations were run in a \hMpc{256 \, (768)} box. 
The  power spectra  of T3,  T4,  and T5  were averaged  over the  four
realisations. 
At the initial  epoch, the 1LPT ICs show a  lack of small-scale power,
which 
 increases when the initial redshift decreases. 
At initial  redshifts of 100 and  50, the initial power  between 1 and
2LPT agree to better than 1\% on all scales, while at initial redshift
of  23,   the  difference  reaches  2\%  at   small  scales  ($\approx
\Mpch{5}$).  
This is expected from Lagrangian perturbation theory, because 1LPT and
2LPT should converge towards high starting redshifts. 
At $z=0$, for the \zi{100}  simulations, the power spectra of the 1LPT
and 2LPT simulations agree to better than 2\% on all scales.  
This agreement is good on scales of $k \le \Mpch{0.2 \, (0.15)}$ for a
starting redshift of 50 (23).
The critical  value of $k$ where  the agreement ceases to  be valid is
independent of the  box size, as shown by the  G simulations, and only
depends on the initial redshift. 
Even when the  initial agreement in the power  spectrum is better than
1\% on all scales, as in the \zi{50} case, the final power spectra can
differ by 2--3\%. 
This  can be  understood because  the  use of  1LPT or  2LPT not  only
affects the initial displacement, but also the initial velocity. 
Therefore, an initial  agreement of 1\% is not  a sufficient condition
to get an accuracy of 1\% at $z=0$ in the power spectrum.

\begin{figure}[t!]
  \begin{center}
    \includegraphics[width=\columnwidth]{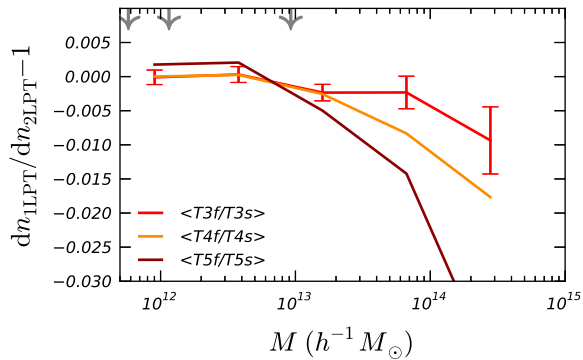}
    \caption{\label{fig:mf_lpt} Influence of the LPT order on
      the halo mass function, same legend as Fig.~\ref{fig:Pk_lpt}.
    }
  \end{center} 
\end{figure}

\begin{figure}[t!]
  \begin{center}
    \includegraphics[width=\columnwidth]{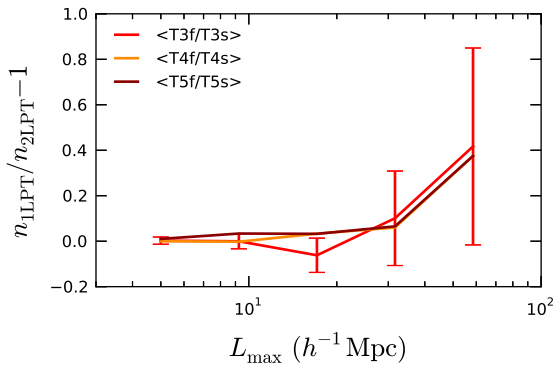}
    \caption{\label{fig:lss_lpt} Influence of the LPT order on
the distribution of the maximal extent of the LSS, same legend 
      as Fig.~\ref{fig:Pk_lpt}. 
    }
  \end{center}
\end{figure}

Figure \ref{fig:mf_lpt}  shows the influence  of the LPT order  on the
mass function  at $z=0$, averaged  over four realisations for  the T3,
T4, and T5 sets of simulations.  
For  the \zi{100} case, the  agreement between 1LPT and 2LPT is within
the error bars. 
At low masses, $M  \lesssim \hMsun{2\times 10^{13}}$, the agreement is
within 1\% in all simulations, but at higher masses, the mass function
is  slightly  lower  in the  case  of  1LPT,  up  to 3\%  at  $M\simeq
\hMsun{10^{14}}$ for the $z_\text{ini} = {23} $ case. 
Figure \ref{fig:lss_lpt}  shows the  effects of the  LPT order  on the
LSS. 
At all  starting redshifts, the  1 and 2LPT simulations  yield similar
results.

\begin{figure}[t!]
  \begin{center}
    \includegraphics[width=222pt]{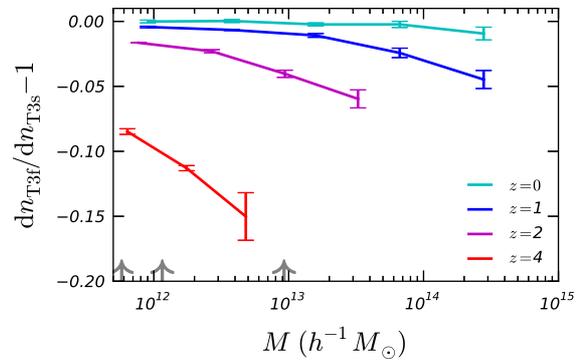}
    \caption{\label{fig:mf_lpt_z}  Evolution of  the 1LPT  versus 2LPT
      mass function for the T4 (initial redshift of 50) simulations.  
    }
  \end{center}
\end{figure}

Here again, it  is interesting to study how  these relations evolve in
time (McBride et al, private communication). 
Figure \ref{fig:mf_lpt_z} shows the time evolution of the ratio of the
mass  functions of the  T3f and  T3s simulations  (\zi{100}), averaged
over 4  realisations, at  $z=0$ (cyan), 1  (blue), 2 (magenta),  and 4
(red).  
At redshifts  $z>0$, the 1LPT always underestimates  the mass function
with respect to 2LPT. 
At $z=0$,  as seen in Fig.~\ref{fig:mf_lpt}, they  are consistent with
each other. 
However, with increasing  redshift, the 1LPT simulation underestimates
the mass function. 
At $z=1$, the underestimation is up to 5\% at $M\simeq\hMsun{10^{14}}$. 
At $z=2$, the underestimation is close  to 5\% for haloes with mass of
\hMsun{10^{13}}, and larger than 1\% at all mass. 
At   $z=4$,   the  underestimation   is   greater   than  10\%   above
\hMsun{10^{12}}. 
If we  want to produce  the halo mass  function on the mass  scale and
below \hMsun{10^{13}} more accurately than 1\%, from 1LPT simulations,
only data after $z=1$, or with more than 50 expansion factors, should
be used.

\subsection{Initial redshift}



In  this section  we  focus on  how  the choice  of starting  redshift
effects the simulation at $z=0$.
This question has been raised in several previous studies
\citep[eg,][]{2007ApJ...671.1160L,                 2009ApJ...698..266K,
  2010ApJ...715..104H, 2013MNRAS.431.1866R}.
However for completeness
we also investigate this issue within our suite of simulations. 
We will  restrict ourselves to  the study of  the 2LPT case,  since we
already focused on the role of the LPT order in the previous section. 
Using too high a starting redshift causes the displacement to be small
with  respect to  the length  resolution, yielding  inaccurate initial
conditions.  
On the  other hand, starting  too late may  break the validity  of the
approximation used for the initial perturbations. 
We reported the rms of the initial displacement in Table \ref{tab:sim}
in units of the mean particle separation.
The  T3 simulations  have  a reasonable  value,  0.26, and  T4 have  a
slightly high value, 0.51.  
The T5 simulations have a  rms displacement of 1.09, which is slightly
larger than the mean separation.  
We note that,  while this value is high, it does  not imply that shell
crossing has occurred, since displacements are correlated. 
On the  other hand, in  the \hMpc{768} box,  T7s (\zi{100}) has  a low
value, 0.089, while those of  T8s (\zi{10}) and T9s (\zi{23}) are more
reasonable, therefore, we  might expect the results of  T8s and T9s to
be more accurate.  
However, we still use T7s as the reference simulation in the large box
in order to be consistent.

\begin{figure}[t!] 
  \begin{center}
      \includegraphics[width=222pt]{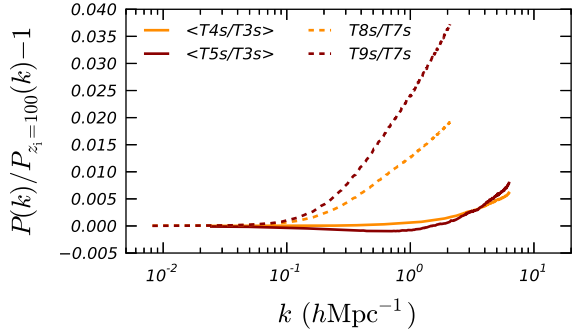}
      \caption{\label{fig:Pk_ini}Influence of  the initial  redshift on
      the power spectrum, 2LPT. 
      The solid lines  show the \hMpc{256} box, averaged  over 4 runs,
      and the dashed lines, the \hMpc{768} one. 
    }
  \end{center} 
\end{figure}

\begin{figure}[t!] 
  \begin{center}
      \includegraphics[width=222pt]{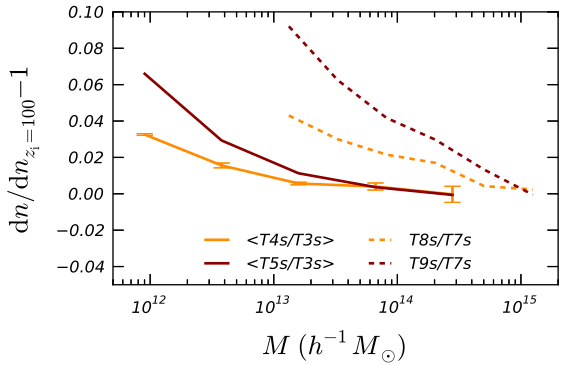}
      \caption{\label{fig:mf_ini}Influence of  the initial  redshift on
      the halo mass function. 
      Same legend as Fig.~\ref{fig:Pk_ini}
    }
  \end{center} 
\end{figure}

Fig.~\ref{fig:Pk_ini} shows  the power spectra of T4s  and T5s divided
by that of T3s, and averaged over 4 realisations. 
Also shown are the power spectra of T8s, and T9s, divided by T7s.  
On large scales, the power  spectra agree exactly, and start to deviate
when approaching the Nyquist frequency.  
For the  \hMpc{256} simulations, the  agreement is better than  1\% on
all scales. 
The  deviation is  slightly larger  in  the \hMpc{768}  box, where  it
reaches 3.5\%. 
However, we must keep in mind that the power spectrum of T7s might not
be accurate, owing to the small initial displacement. 
Figure \ref{fig:mf_ini} shows the  effects of the starting redshift on
the halo mass function at $z=0$.
Shown are  the mass functions of T4s  and T5s divided by  that of T3s,
and T8s, and T9s, divided by that of T7s.  
On large  masses, in  the small box  case, all simulations  agree, but
when going  toward low masses ($<\hMsun{10^{13}}$), a  clear effect of
the starting redshift can be seen.  
Simulations with a lower starting redshift overestimate the number
density of low-mass haloes by up to 4\% (T4s) and 8\% (T5s).  
For the larger box, the results is even more striking.
At  $M\simeq \hMsun{10^{15}}$,  the  three simulations  agree, but  at
lower masses the differences appear, 
up to 10\% at $M\simeq \hMsun{10^{13}}$ for the T9s case.

\begin{figure}[t!] 
  \begin{center}
    \includegraphics[width=222pt]{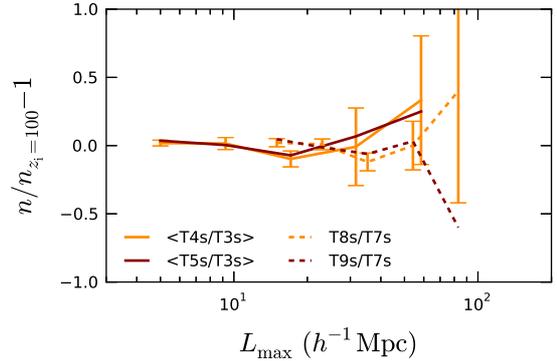}
    \caption{\label{fig:lss_ini} Influence of  the initial redshift on
      the maximal extent of LSS, same legend as \ref{fig:Pk_ini}.  
    }
  \end{center}
\end{figure}


Figure \ref{fig:lss_ini} shows the effects of the starting redshift on
the distribution of the LSS. 
At the small-size end of the distribution, all simulations agree very well. 
On larger sizes, the uncertainty becomes more important.
To the accuracy reached  by the simulations, no significant difference
can be seen between the different simulations. 
If we combine this result with that in the previous section, it can 
be said that  the size distribution of LSS  estimated from simulations
with  a relatively  low  initial  redshift and  1LPT  ICs is  reliable
(i.e. \citealt{2012ApJ...759L...7P} who used \zi{32}, 1LPT, and a mean
particle separation of \hMpc{1.2}).

\subsection{$N$-body code comparison: \go{} versus \gad{-3}}
\label{sec:nb}
 
\begin{figure}[t!]
  \begin{center}
      \includegraphics[width=222pt]{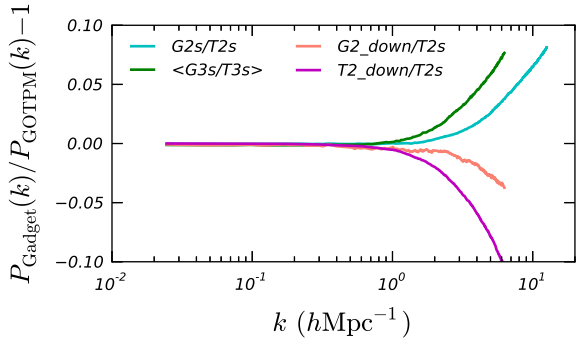}
       \caption{\label{fig:Pk_GT}Influence of the $N$body code on
      the power spectrum.
      Shown are the power spectra of \gad{} simulations divided by the
      corresponding \go{} one. 
      The G3s/T3s results were averaged over the four runs. 
      A downsampled version of  the G2s and T2s simulations, T2s\_down
      and G2s\_down, are also shown in magenta and salmon.
    }
  \end{center}
\end{figure}

\begin{figure}[t!]
  \begin{center}
    \includegraphics[width=222pt]{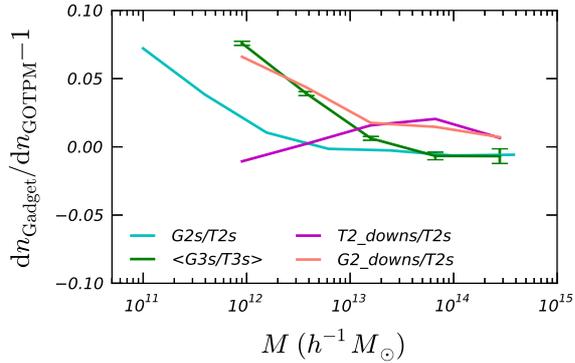}
    \caption{ \label{fig:mf_GT}Influence of the $N$body code on
      the halo mass function.
      Same legend as Fig. \ref{fig:Pk_GT}
    }
  \end{center}
\end{figure}

In this section, we compare the two cosmological codes.
In order to test the  consistency between \go{} and \gad{}, some
simulations, namely  T2s, T3s, and the  first run of T4s  (T4s Run 1),
were also run using \gad{-3}.  

Figure  \ref{fig:Pk_GT}   shows  the  power  spectra   of  the  \gad{}
simulations divided by those of the corresponding \go{} ones. 
On large scales, the \gad{-3} and \go{} simulations agree very well. 
However,  on  small scales  $k>k_\text{Ny}/4$,  the \gad{-3}  simulation
start to  show an  excess of power  with respect to  the corresponding
\go{} simulation, up to 8\% at the Nyquist frequency. 
This  effect  is  more  important  than the  effects  of  the  starting
redshift. 
To better understand this difference, we also run a downsampled version
of the  T2s simulation with $512^3$ particles  using \go{} (T2s\_down)
and \gad{} (G2s\_down). 
In these  simulations, the large-scale  initial power is the  same as
the  T2s simulation,  which has  1024$^3$ particles,  but  the initial
density field has been binned  down to $512^3$, losing the small-scale
power information.  
The ratio  of T2s\_down and G2s\_down  to T2s are shown  in salmon and
magenta. 
The  large-scale power  agree  exactly, as  expected, and  differences
occur on small scales. 
T2s\_down shows some lack of small-scale power at $k>\Mpch{1}$. 
In the case of G2s\_down, the lack of small-scale power is balanced by
the extra power of the \gad{} runs, yielding a smaller difference. 

\begin{figure}[t!]
  \begin{center}
    \includegraphics[width=222pt]{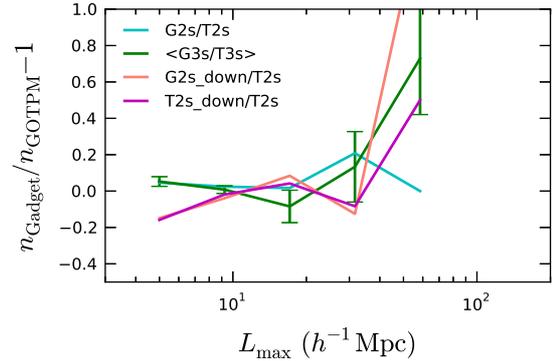}
    \caption{\label{fig:lss_GT} Influence of  the $N$-body code on the
      size distribution of the large-scale structures. 
      Same legend as Fig. \ref{fig:Pk_GT} 
    }
  \end{center}
\end{figure}

Figure \ref{fig:mf_GT} shows the ratio  of the mass function of \gad{}
and \go{} simulations
The agreement is very good, with 1\% accuracy at high masses.
At  fixed resolution,  at high  masses,  the agreement  is very  good,
better than 1\%. 
However, at  low masses, $M<\hMsun{10^{13}}$,  \gad{} overestimates the
mass function with respect to \go, up to 8\%. 
The two downsampled simulations show an interesting feature. 
The  high-mass end of  the mass  function ($M>\hMsun{10^{13}}$)  is 2\%
higher than the T2s one. 
However, on low masses, they show different behaviours. 
G3s\_down  shows the same  characteristics as  the other  \gad{} runs,
with an  excess of  low-mass haloes compared  to T2s,  while T2s\_down
shows closer results to T2s.  
It  should be noted  that G2s,  the higher-resolution  simulation with
\gad{}, agrees with T2s down to \hMsun{10^{12}} to better than 2\%. 
The   fact  that   T2\_down   approaches  both   T2s   and  G2s   near
\hMsun{10^{12}} while  G2s\_down deviates from them,  means that \gad{}
is overproducing the haloes on small-mass scales.  

Finally, Fig.  \ref{fig:lss_GT} shows the effect of  the $N$-body code
on the size distribution of the LSS. 
 No clear effect can be seen beyond the fluctuations.

\section{Discussion and conclusion}

\label{sec:ccl}
We  have created  a suite  of $N$-body  simulations, run  with various
initial conditions that explored changes in the LPT order, the initial
redshift, and the pre-initial configuration.  
This  simulation  dataset was  used  to  quantify  the sensitivity  of
observable statistics,  such as the  density power spectrum,  the halo
mass  function,  and  the  distribution  of the  LSS  extent,  on  the
numerical parameters.  
Our main findings are:
\begin{itemize}
\item 
We found that the choice of the pre-initial conditions does not affect
the simulation results significantly: the difference is less than 1\%
for the power spectrum and the halo mass function. 
However, it  should be  pointed out that  the density  fluctuations are
spuriously very  high on small scales  (up to twice  the mean particle
separation) due to the initial white noise power.
Even  though this  power  does not  grow  gravitationally, the  excess
persists until it is exceeded by the growing ``true'' power.  
This  manifests itself  most  significantly at  higher redshift  where
numerical artefacts may be confused with small-scale physics.  
At $z=3$, we  found a lack of small-scale  power and small-mass haloes
in the glass simulations.
However,  after  $z=2$  we   found  no  statistical  evidence  for  a
systematic difference between the halo mass functions of the glass and
mesh preIC simulations.

\item 
One  drawback in glass  preICs is  that choosing  too high  a starting
redshift  introduces  larger errors  compared  to  the lower  starting
redshift case. 
This may be because of the difficulty in realising the small-amplitude
density  fluctuations by a  particle distribution  with a  white noise
power. 
The  interpolation of  the initial  displacement field  from  the mesh
points to particle positions can be another source of errors.

\item 
We confirmed that the 1LPT  underestimates the power spectrum on small
scales  compared to  the  2LPT simulations  and  that the  differences
increase as the starting redshift decreases.  
The halo mass function is  slightly underestimated on the highest mass
scales probed ($\simeq\hMsun{10^{14}}$) in the 1LPT case. 
The  difference   is  approximately   1\%  near  the   mass  scale  of
\hMsun{10^{14}}  when  the starting  redshift  is  50,  and the  trend
increases for lower starting redshifts.
We find no statistically significant effect on the order of the LPT on
the size distributions of LSS.

\item We  found that the underestimation  in the mass  function in the
  1LPT simulations  increases at high redshift.
  For an initial redshift of 100, it  is less than 1\% at $z=0$ and 1,
   about  5\% at  $z=2$, and more  than 20\%  at $z=4$ on  the mass
  scale of \hMsun{10^{13}}.

\item 
We found that the starting redshift makes a systematic and
statistically  significant impact  on the  FOF halo  mass  function at
small masses ($<\hMsun{10^{13}}$).
The  choice  of  relatively  low  starting redshifts  ($z=50$  and  23
compared to 100) yielded an  over-estimation of the mass function by a
factor of more than 5\% near the mass scale of \hMsun{10^{12}}.
We do  not find  significant differences  in the PS and the  LSS size
distribution for different starting redshifts.

\item We compared the gravitational N-body integration codes \go{} and
  \gad{} on cosmologically significant scales. 
  Both  codes yield  similar large-scale  power spectra  and high-mass
  mass functions, but  it is found that \gad{}  overproduces the small
  mass  haloes  on  the  mass  scale  below  about  $10^3$  simulation
  particles.  
  The  overproduction reaches  about  8\%  at a  mass  scale of  $100$
  particles.

\end{itemize}

Our    results     using    2LPT    agree     with    previous    studies
\citep{2010ApJ...715..104H, 2013MNRAS.431.1866R}, and here we stressed
again the need for the use of 2LPT.  
We agree with \citet{2013MNRAS.431.1866R} about the need for 2LPT ICs,
especially if  one is interested  in the high-redshift  ($z\gtrsim 3$)
mass function, where 1LPT ICs underestimate it more seriously. 
However, \citet{2013MNRAS.431.1866R} warned about false convergence in 
high starting redshift simulations. 
Before  running  $N$-body simulations,  such  tests  should always  be
performed.  
In addition  to the  power spectrum and  the halo mass  function, which
have been  extensively studied  before, the use  of a  new statistics,
namely the size distribution of  LSSs, provides us with information on
the large scale.

In this study we have ignored hydrodynamical effects. 
While the large-scale evolution is predominately driven by gravity, on
smaller  scales the role  of the  baryons plays  a more  important but
complicated          role          \citep[e.g.][]{2006ApJ...640L.119J,
  2008ApJ...672...19R,         2011MNRAS.415.3649V,2012MNRAS.423.2279C,
  2013JKAS...46....1H}.  
\citet{2012MNRAS.423.2279C}  showed that, even  in a  purely adiabatic
case (i.e., no cooling nor star formation), the halo mass function may
change  up to $\approx  7\%$ for  an overdensity  of 500,  showing the
impact of the baryons in the inner parts of the haloes.  
In a next  paper, we will extend the  statistical tests presented here
to cosmological hydrodynamical simulations.

{
Here  we chose  to study  the  convergence of  simulation results  for
different numerical setups.
Once simulations  with a given  cosmolgy can achieve 1\%  precision in
the mass function, it becomes possible to test different cosmologies.  
Studying the cosmology dependence of  the mass function is also a very
important topic.  
For  instance,   \citet{2011MNRAS.410.1911C}  showed  that   the  mass
function is  close to universal  for cosmologies with  close expansion
histories, but deviates from  universality for other cosmologies (Dark
Energy).

Figure \ref{fig:mf_lpt_z} shows  that there is a few percent difference in
the mass function between 1LPT and 2LPT at each redshift interval. 
The effects of other simulation setups on the mass function are smaller.
On the other hand,  \citet{2014ApJ...780...34L} find the mass function
evolves by a factor of a few across similar redshift intervals.  
Therefore, the effect of evolution dominates the changes in
the amplitude and shape of the mass function. At a given redshift, the
uncertainties in  the current observational  data are still  too large
compared to  the numerical uncertainties found in  our paper. However,
when the  size of observations is  increased, it will  be necessary to
have an accurate  theoretical mass function that can  be compared with
the  more accurately  determined observed  mass function  in  order to
constrain cosmology and astrophysics. 
}

\section*{Acknowledgements}
We thank KIAS Center  for Advanced Computation for providing computing
resources. 
We  thank  \modif{the referee,  Fabio  Governato,  for his  comments},
Volker Springel for providing us with \gad{-3}, and
Cristiano Sabiu for his comments on the paper.

\bibliographystyle{aa} 
\bibliography{cosmoIC}

\appendix
\section{Power spectrum estimation}
\label{sec:Pk}

\begin{figure}[t!]
  \begin{center}
    \includegraphics[width=222pt]{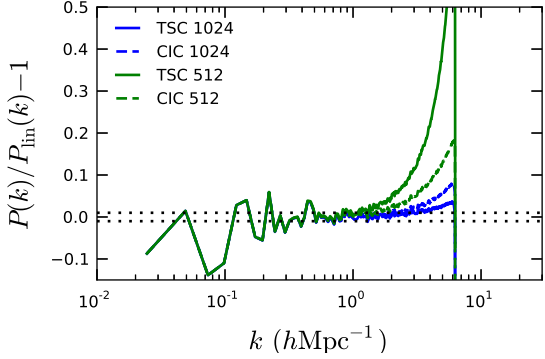}
    \caption{\label{fig:Pk}Power spectrum of a simulation (T3s) at the
      initial epoch  $z = 100$,  with $N_\text{grid} = $  512 (green),
      and 1024 (blue); computed with the CIC (dashed), and TSC (solid)
      schemes} 
  \end{center} 
\end{figure}

Our  code for computing  the power  spectrum is  based on  the fastest
Fourier transform in the west (FFTW) library.
The density is evaluated on  a user-defined grid (we chose $N_\text{g}
= 2^3 N_\text{p}$),  using either a nearest grid  point (NGP), CIC, or
TSC assignment scheme \citep{1981csup.book.....H}.
The  modes are  binned  using a  cloud-in-cell  scheme assignment  to
attenuate  the large-scales fluctuations  due to  the small  number of
modes in the first bins.
The                  code                  is                 publicly
available\footnote{\url{http://aramis.obspm.fr/~lhuillier/codes.php}}, is 
MPI-parallel, 
is able to read \gad{-3} and \go{} snapshot files, and has been tested 
using up to a 4096$^3$ grid on 96 MPI tasks. 
Figure \ref{fig:Pk}  shows the initial power spectrum  of the T3s
simulation, normalised by the input linear power spectrum, and computed
on  several grid  sizes  and  using different  schemes,  that are  the
cloud-in-cell (CIC,  dashed lines),  and triangular shape  cloud (TSC,
solid lines)  mass scheme  assignment, using a  grid with a  number of
cells per dimension of 1024 (blue), and 512 (green).  
At fixed  resolution, the  TSC yields more  accurate results  than the
CIC.
The TSC, 1024 yields a measurement to 2\% accuracy on all scales. 
This setting corresponds to with $N_\text{grid}^3 = 2^3 N_\text{p}^3$,
which we will keep for all  measurements of the power spectrum in this
paper.  
Note  that  \citet{2010ApJ...715..104H} also  used  the same  setting,
however using a CIC scheme. 
 
\section{Effect of the halo-finder}
\begin{figure}[h] 
  \begin{center}
    \includegraphics[width=222pt]{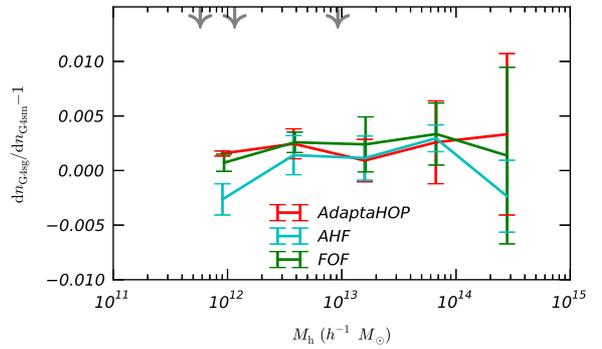}
    \caption{\label{fig:ahop}Comparison of the FOF, AdaptaHOP, and AHF
      mass functions for the G4sm anfd G3sg simulations.} 
  \end{center}
\end{figure}

{
It is  beyond the scope  of this article  to study the  convergence or
disagreements among different  halo finders.
 \citet{2011MNRAS.415.2293K} performed such a comparison.
In figure \ref{fig:ahop}, we compare the mass function of the G4sm and
G4sg simulations calculated by our FOF algorithm, AdaptaHOP
\citep{2004MNRAS.352..376A,2009A&A...506..647T}   and   AMIGA's   halo
finder \citep[AHF,][]{2004MNRAS.351..399G, 2009ApJS..182..608K}. 
AdaptaHOP and AHF are subhalo finders, but the subhalo detection
can  be omitted  and we  only  consider the  main haloes  in order  to
compare to the FOF haloes.  

From Fig.~\ref{fig:ahop}, it is clear that,
although  the mass  functions of  FOF, AHF,  and AdaptaHOP  haloes are
slightly different, the halo finder does not play an important role.
The  three halo finders  agree within  the error  bars, except  in the
smallest mass bin for AHF, and the agreement is below the 1\% level.
We do not  see systematic differences compared to the  FOF case, so we
conclude that this is rather insensitive to the halo finder.
}

\end{document}